%% file: main.tex
\documentclass[a4paper,11pt]{article}
\usepackage{pos}
\usepackage{subcaption}
\usepackage{booktabs}

\newcommand{\atel}[1]{\href{http://www.astronomerstelegram.org/?read=#1}{ATel {#1}}}

\title{IceCube search for neutrinos from novae}

\ShortTitle{IceCube search for neutrinos from novae}

\author{The IceCube Collaboration \\{\normalsize \normalfont(a complete list of authors can be found at the end of the proceedings)}\\}

\emailAdd{jessie.thwaites@icecube.wisc.edu}
\emailAdd{justin.vandenbroucke@wisc.edu}

\abstract{
Despite being one of the longest known classes of astrophysical transients, novae continue to present modern surprises. The \textit{Fermi}-LAT discovered that many if not all novae are GeV gamma ray sources, even though theoretical models had not even considered them as a possible source class. More recently, MAGIC and H.E.S.S. detected TeV gamma rays from a nova. Moreover, there is strong evidence that the gamma rays are produced hadronically, and that the long-studied optical emission by novae is also shock-powered. If this is true, novae should emit a neutrino signal correlated with their gamma-ray and optical signals. We present the first search for neutrinos from novae. Because the neutrino energy spectrum is expected to match the gamma-ray spectrum, we use an IceCube DeepCore event selection focused on GeV-TeV neutrinos. We present results from two searches, one for neutrinos correlated with gamma-ray emission and one for neutrinos correlated with optical emission. The event selection presented here is promising for additional astrophysical transients including gamma-ray bursts and gravitational wave sources.

\vspace{4mm}
{\bfseries Corresponding authors:}
Jessie Thwaites$^{1}$, Justin Vandenbroucke$^{1*}$\\
{$^{1}$ \itshape University of Wisconsin-Madison and Wisconsin IceCube Particle Astrophysics Center}\\[4mm]
$^*$ Presenter

\ConferenceLogo{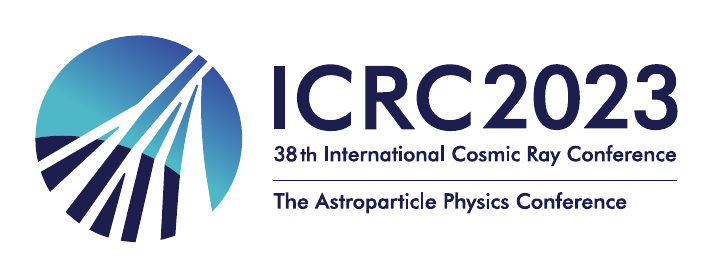}

\FullConference{The 38th International Cosmic Ray Conference (ICRC2023)\\ 26 July -- 3 August, 2023\\ Nagoya, Japan}
}

\begin{document}

\maketitle

\section{Introduction}

Novae, one of the longest known, historical classes of astrophysical transients, continue to surprise us.  In addition to the GeV gamma rays discovered by Fermi LAT, MAGIC and H.E.S.S. have recently discovered nova emission approaching the TeV scale.  Those novae that are brightest in the optical band have been detected in gamma rays, indicating that many or potentially all novae emit gamma rays and that those that have not been detected are simply below instrument detection thresholds.  Furthermore, strong correlation between the optical and gamma-ray light curves indicates that shock acceleration powers not only the gamma-ray emission, but also the optical emission and therefore the total bolometric power of novae.  If the gamma-ray emission mechanism is hadronic rather than leptonic, neutrinos are expected to accompany the gamma-ray signal.  Novae could even be brighter in neutrinos than in gamma rays if the gamma rays are partially absorbed within the source.  We present the first search for neutrinos from novae, using the IceCube Neutrino Observatory. We use a new event selection, GRECO Astronomy \cite[Appendix B]{IceCube:2022lnv}, which uses IceCube-DeepCore, the densely instrumented inner subarray of IceCube. 

\subsection{Nova Sample}
\begin{figure*} 
    \centering
    \includegraphics[width=0.8\textwidth]{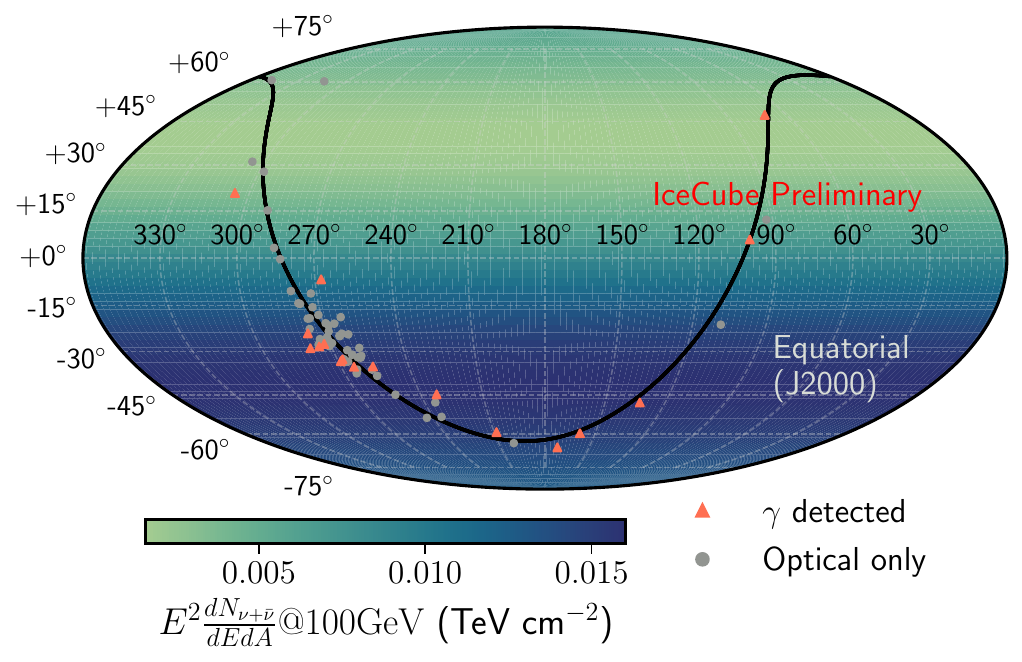}
    \caption{Locations of novae considered in this analysis. Gamma-ray detected novae are shown in orange, including nova RS Oph, while those only detected at other wavelengths are shown in gray. These novae are shown on top of the sensitivity of the IceCube analysis to a single nova with a time window of one day. }
    \label{fig:skymap}
\end{figure*}

The catalog of novae used in this search is described in full in \cite[Appendix A]{IceCube:2022lnv}. At the time of that study, we only had processed GRECO Astronomy data spanning April 2012 to May 2020. With the addition of two more years of processed data (up to October 2022), it is now possible to include nova RS Ophiuchi in the list of neutrino source candiates. All of these novae, including RS Oph, are shown in Figure \ref{fig:skymap}.

\subsection{Analysis techniques}
The analysis uses an unbinned maximum likelihood technique described in \cite{IceCube:2022lnv}. We assume neutrino emission with a power-law spectrum $dN_{\nu+\bar{\nu}}/dE \propto E^{-\gamma}$, with $\gamma$ the spectral index. We maximize the likelihood and define our test statistic (TS) as the log-likelihood ratio between the best-fit signal and the background hypothesis
\begin{equation}
    \mathrm{TS} = - 2 \ln \left[\frac{\mathcal{L}(\hat{n}_{s}, \hat{\gamma})}{\mathcal{L}(n_{s}=0)}\right],
\end{equation}
with $\hat{n}_s$ the best-fit number of signal events and $\hat{\gamma}$ the spectral index of a given source. In the case where $\hat{n}_{s}=0$, our observed $\mathrm{TS}=0$, which represents an underfluctuation with respect to the expected number of background events in the short time window considered.

\section{Individual gamma-ray correlation analysis} 
\label{sec:individual}
We search for neutrino emission coincident with each of the gamma-ray detected novae, as described in \cite{IceCube:2022lnv}. For each nova, we define our on-time window to be the detection time of the nova by \textit{Fermi}-LAT. The time windows are chosen to match those given by \cite{Gordon:2020fqv} for most novae, with a few exceptions. Two novae, V745 Sco and V1535 Sco, were identified as candidate gamma-ray sources  because they did not reach a detection at the 3$\sigma$ level in \cite{Franckowiak:2017iwj}, so we use the detection times reported in that paper. Two other novae, V3890 Sgr and V1707 Sco, had more recent outbursts reported by \textit{Fermi}-LAT in real-time via the Astronomer's Telegram (ATel), so we use their detection times as reported in their ATels (\atel{13114} and \atel{13116}, respectively). The time window for each nova is given in Table \ref{tab:gamma_results}.

\begin{figure*} 
    \centering
    \includegraphics[width=.98\textwidth]{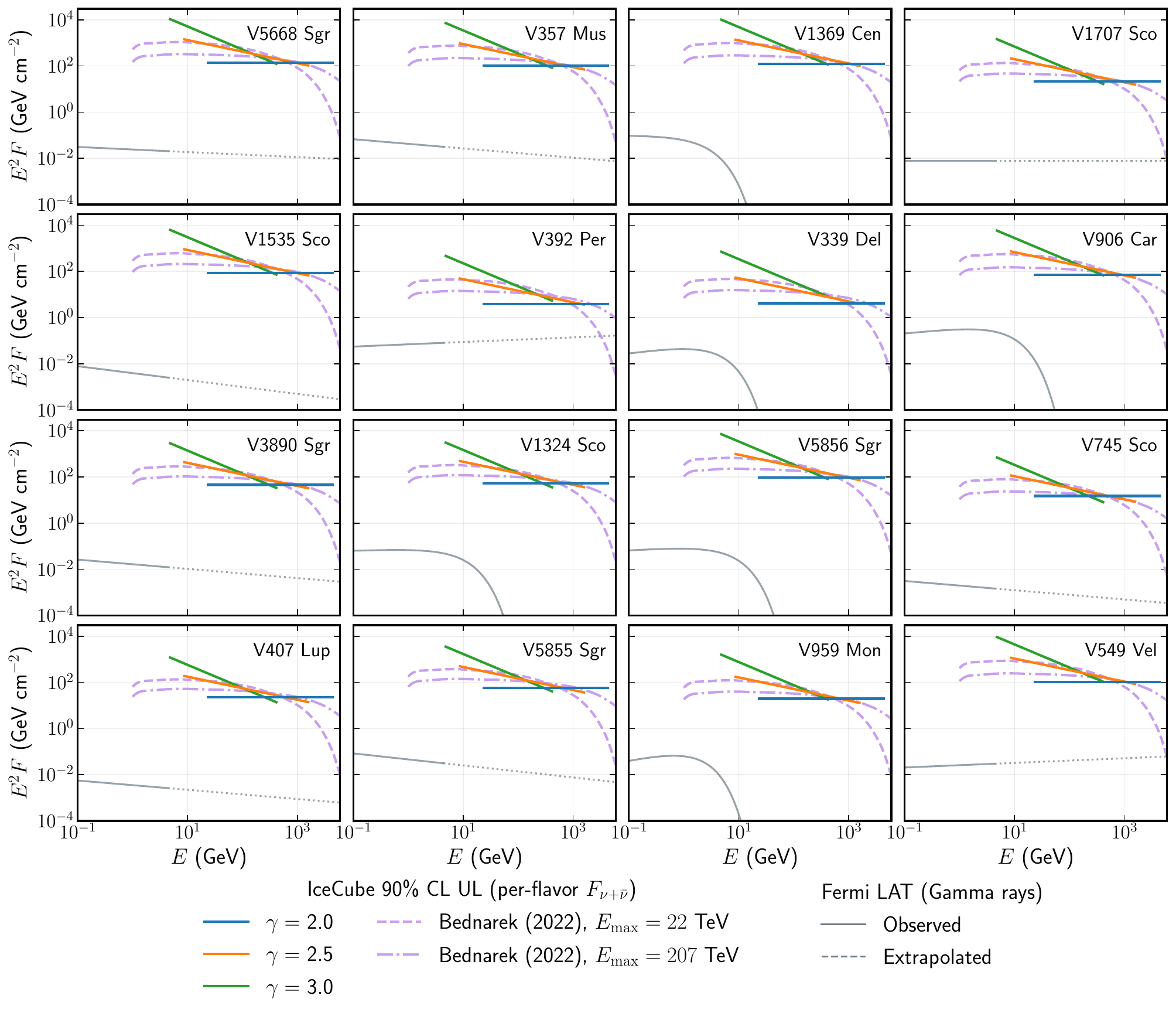}
    \caption{Upper limits on neutrino emission from all of the novae analyzed in the gamma-ray correlation analysis (adapted from \cite{IceCube:2022lnv}). Upper limits on power laws span the central 90\% energy ranges of the GRECO Astronomy data set for the given spectral index, and are rescaled by the systematic uncertanties to provide conservative upper limits, as discussed in \cite[Section 4.3]{IceCube:2022lnv}. In addition to constraining power-laws, we also inject the spectral shapes from models in~\cite{Bednarek:2022vey}, and our upper limits on those spectra are shown in pink. We compare these fluxes to the \textit{Fermi}-LAT measured gamma-ray fluxes (gray). For those gamma-ray detected novae which did not show evidence for a cutoff in the gamma-ray spectra, we show the lines as dotted past the global cutoff energy found in~\cite{Franckowiak:2017iwj}.}
    \label{fig:ul_panel}
\end{figure*}

We do not detect any significant neutrino emission from any of the novae, so we set 90\% confidence level (CL) upper limits. All of these limits are plotted in Fig. \ref{fig:ul_panel}. 

\subsection{Search for neutrinos from RS Ophiuchi}
For the analysis searching for neutrino emission from nova RS Oph, we use two additional processed years of GRECO Astronomy data, including data up to October 2022. We use the same analysis as that of the other 16 gamma-ray detected novae. To chose the time window considered for RS Oph, we again use the detection time window by \textit{Fermi}-LAT given in \cite{MAGIC:2022rmr}, with a total duration of 30 days. This also covers the TeV detections of RS Ophiuchi reported by H.E.S.S. \cite{HESS:2022qap} and MAGIC \cite{MAGIC:2022rmr}. 

\begin{figure*} 
    \centering
    \includegraphics[width=0.7\textwidth]{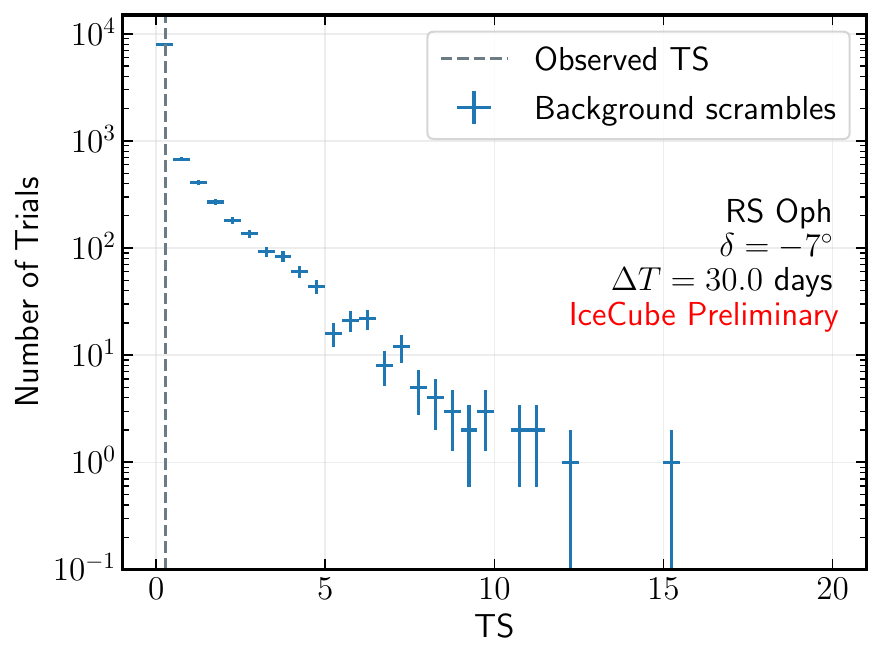}
    \caption{Background test statistic distribution for an analysis of GRECO data using a 30 day time window at the location of nova RS Oph, with the unblinded TS for nova RS Oph shown in the grey dashed line. }
    \label{fig:tsd}
\end{figure*}

\begin{figure*} 
    \centering
    \includegraphics[width=0.7\textwidth]{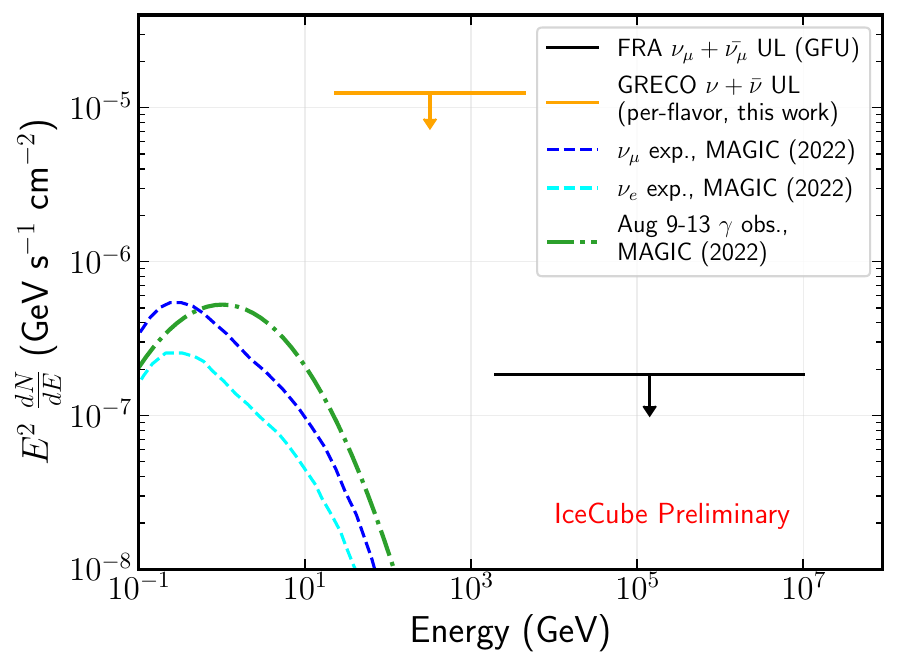}
    \caption{Upper limits for GRECO and Fast Response (\atel{14851}), which uses the GFU sample for nova RS Oph. In addition, the Log-Parabola fit to the MAGIC and Fermi-LAT observations from the outburst (Aug 9-13) is plotted in the green dash-dotted curve \cite[Suppl. Table 3]{MAGIC:2022rmr}. Also shown are neutrino expectations calculated by \cite[Suppl. Fig. 1]{MAGIC:2022rmr} assuming p-p interactions. }
    \label{fig:magic_comp}
\end{figure*}

\begin{figure*} 
    \centering
    \includegraphics[width=0.9\textwidth]{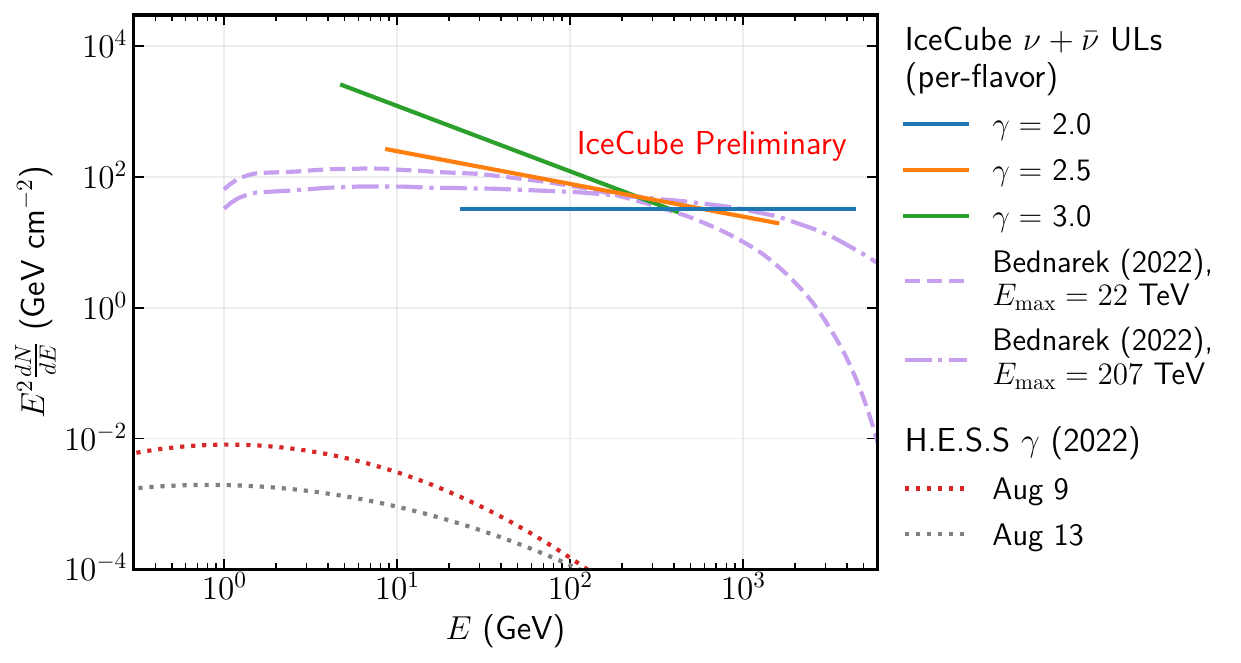}
    \caption{Upper limits on neutrino emission from RS Ophiuchi. Upper limits on power laws span the central 90\% energy ranges of the GRECO Astronomy data set for the given spectral index, and account for systematic uncertainties as discussed in \cite[Sec. 4.3]{IceCube:2022lnv}. In addition to constraining power-laws, we also inject the spectral shapes from models in \cite{Bednarek:2022vey}, and our upper limits on those spectra are shown in pink. Joint fits to the H.E.S.S. and \textit{Fermi}-LAT observations of RS Ophiuchi for August 9th and August 13th are shown as dotted curves \cite[Fig.3]{HESS:2022qap}. }
    \label{fig:ul_panel_rs_oph}
\end{figure*}

No significant neutrino emission is found from nova RS Oph. We compare the observed test statistic (TS) to a distribution of 10,000 pseudo-experiments, generated from scrambling our data in right ascension. The observed test statistic, compared to the background TS distribution for this source can be seen in Fig \ref{fig:tsd}. We then place 90\% CL upper limits using the one-sided Neyman-Pearson construction. The upper limit assuming a spectrum $dN_{\nu+\bar{\nu}}/dE \propto E^{-2}$ is plotted in Fig. \ref{fig:magic_comp}, compared to the upper limit for high-energy neutrino emission computed in real-time using the Fast Response Analysis \cite{IceCube:2020mzw}, which uses the Gamma-ray follow-up (GFU, \cite{IceCube:2016cqr}) data sample, with results reported in \atel{14851}. These are compared to a joint fit for gamma-ray observations from \textit{Fermi}-LAT and MAGIC and neutrino expectations calculated in \cite{MAGIC:2022rmr}. 

We set upper limits for a power-law spectrum with 3 different spectral indices, as well as a physical model for neutrino emission in novae \cite{Bednarek:2022vey}, shown in Fig. \ref{fig:ul_panel_rs_oph}. We also show a joint fit to the gamma-ray observations from \textit{Fermi}-LAT and H.E.S.S. at the peak of the gamma-ray emission (August 9th, 2021), and observations five days later (August 13th, 2021). An updated table of results, including analysis results for nova RS Oph, is provided in Table \ref{tab:gamma_results}. 

\begin{table*}
\centering
    \tabcolsep=4.5pt\relax
    \small
\begin{tabular}{ccccccccccc}
\toprule
Name &                 $\alpha$ & $\delta$ & MJD$_{\mathrm{start}}$ & MJD$_{\mathrm{stop}}$ &  $\Delta$T & TS & $\hat{n}_{s}$ &$\hat{\gamma}$ &   pre-trial & $E^2F_{\nu+\bar{\nu}}$ @ 1~TeV \\
&  &  & &  &(days)  &  & & &   $p$-value & (GeV cm$^{-2}$)\\
\midrule
V1324 Sco & 267.7$^{\circ}$ & -32.6$^{\circ}$ &                56093.0 &               56110.0 & 17.0 &         0.00 &                     0.0& -- & 1.000 & 52.9\\
 V959 Mon &   99.9$^{\circ}$ & +5.9$^{\circ}$ &                56097.0 &               56119.0 & 22.0 &         1.14 &                  14.8& 2.75 & 0.197 & 19.5\\
 V339 Del & 305.9$^{\circ}$& +20.8$^{\circ}$ &                56520.0 &               56547.0 & 27.0 &         0.00 &                     0.0& -- & 1.000 & 4.21\\
V1369 Cen & 208.7$^{\circ}$& -59.2$^{\circ}$ &                56634.0 &               56672.0 & 38.0 &          0.01 &                   3.8& 4.00 & 0.070 & 125.9\\
 V745 Sco & 268.8$^{\circ}$& -33.2$^{\circ}$ &                56694.0 &               56695.0 & 1.0 &         0.00 &                     0.0& -- & 1.000 & 14.9\\
V1535 Sco & 255.9$^{\circ}$& -35.1$^{\circ}$ &                57064.0 &               57071.0 & 7.0 &         6.95 &                  59.9& 3.02 & 0.002 & 85.2\\
V5668 Sgr & 279.2$^{\circ}$& -28.9$^{\circ}$ &                57105.0 &               57158.0 &  53.0 &        1.30 &                  62.0& 3.16 & 0.112 & 139.2\\
 V407 Lup & 232.3$^{\circ}$& -44.8$^{\circ}$ &                57657.0 &               57660.0 & 3.0 &         0.00 &                     0.0& -- & 1.000 & 22.4\\
V5855 Sgr & 272.6$^{\circ}$& -27.5$^{\circ}$ &                57686.0 &               57712.0 & 26.0 &           0.00 &                     0.0& -- & 1.000 & 56.6\\
V5856 Sgr & 275.2$^{\circ}$& -28.4$^{\circ}$ &                57700.0 &               57715.0 & 15.0 &          5.13 &                  40.3& 2.81 & 0.015 & 94.0\\
 V549 Vel & 132.6$^{\circ}$& -47.8$^{\circ}$ &                58037.0 &               58070.0 & 33.0 &          0.22 &                  22.4& 4.00 & 0.062 & 103.2\\
 V357 Mus & 171.6$^{\circ}$& -65.5$^{\circ}$ &                58129.0 &               58156.0 & 27.0 &          0.01 &                   5.0& 4.00 & 0.115 & 104.6\\
 V906 Car & 159.1$^{\circ}$& -59.6$^{\circ}$ &                58216.0 &               58239.0 & 23.0 &         0.00 &                     0.0& -- & 1.000 & 73.0\\
 V392 Per &  70.8$^{\circ}$& +47.4$^{\circ}$ &                58238.0 &               58246.0 & 8.0 &          0.88 &                  15.4& 3.64 & 0.373 & 3.84\\
V3890 Sgr & 277.7$^{\circ}$& -24.0$^{\circ}$ &                58718.0 &               58739.0 & 21.0 &          0.00 &                     0.0& -- & 1.000 & 45.1\\
V1707 Sco & 264.3$^{\circ}$& -35.2$^{\circ}$ &                58740.0 &               58744.0 & 4.0 &          0.00 &                     0.0& -- & 1.000 & 21.2\\
RS Oph & 267.6$^{\circ}$ & -6.7$^{\circ}$ & 59434.8 & 59464.8 & 30.0 & 0.28 & 3.8 & 2.4 & 0.255 & 32.2 \\
\bottomrule
\end{tabular}
\caption{Results from the gamma-ray correlation analysis. MJD$_{\mathrm{start}}$ and MJD$_{\mathrm{stop}}$ represent the beginning and end of when a nova was detected by \textit{Fermi}-LAT, respectively. This is the same time window used for the neutrino search. The observed test statistic (TS), best-fit number of signal events ($\hat{n}_{s}$) and best-fit spectral index ($\hat{\gamma}$) for the neutrino search from each nova are given here (Note that $\hat{\gamma}$ is undefinded in the case of $\hat{n}_{s}=0$). The $p$-values shown are before accounting for the factor accrued from performing multiple searches. The final column gives the time-integrated flux upper limit for each nova, assuming a power-law spectrum with a spectral index of 2.0 ($dN_{\nu+\bar{\nu}}/dE \propto E^{-2}$) and including the systematic uncertainty as discussed in \cite[Section 4.3]{IceCube:2022lnv}.}
\label{tab:gamma_results}
\end{table*}

\section{Stacking analysis}

\begin{figure*}
    \centering
    \includegraphics[width=0.46\textwidth]{ 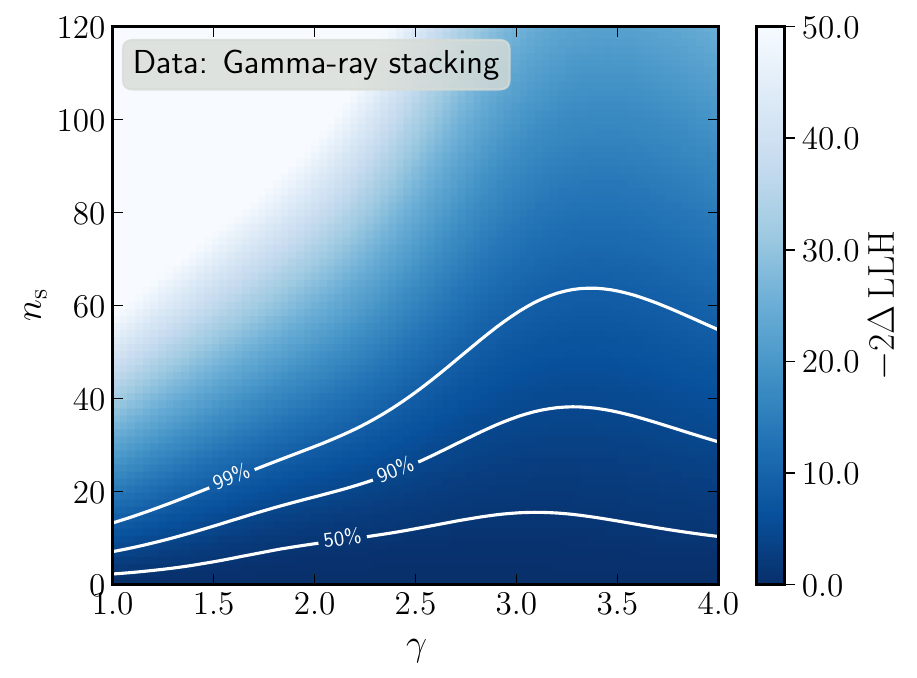}
    \includegraphics[width=0.46\textwidth]{ 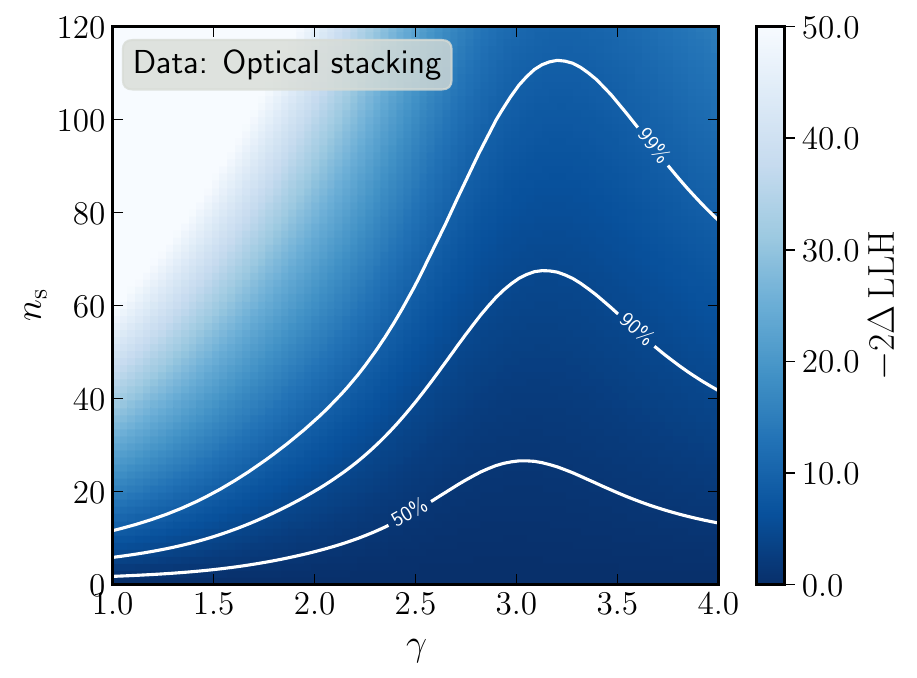}
    \includegraphics[width=0.46\textwidth]{ 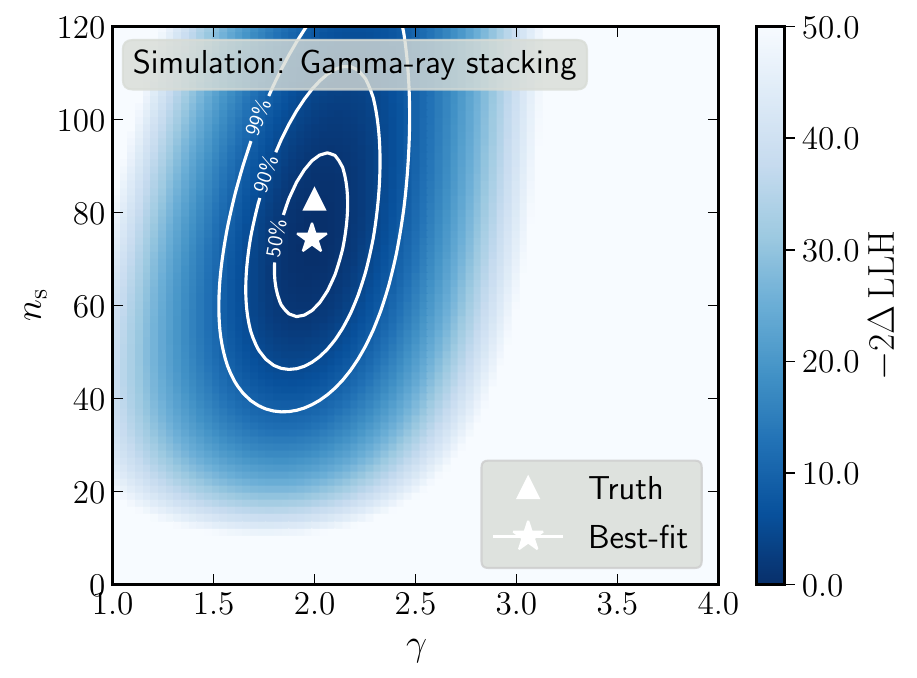}
    \includegraphics[width=0.46\textwidth]{ 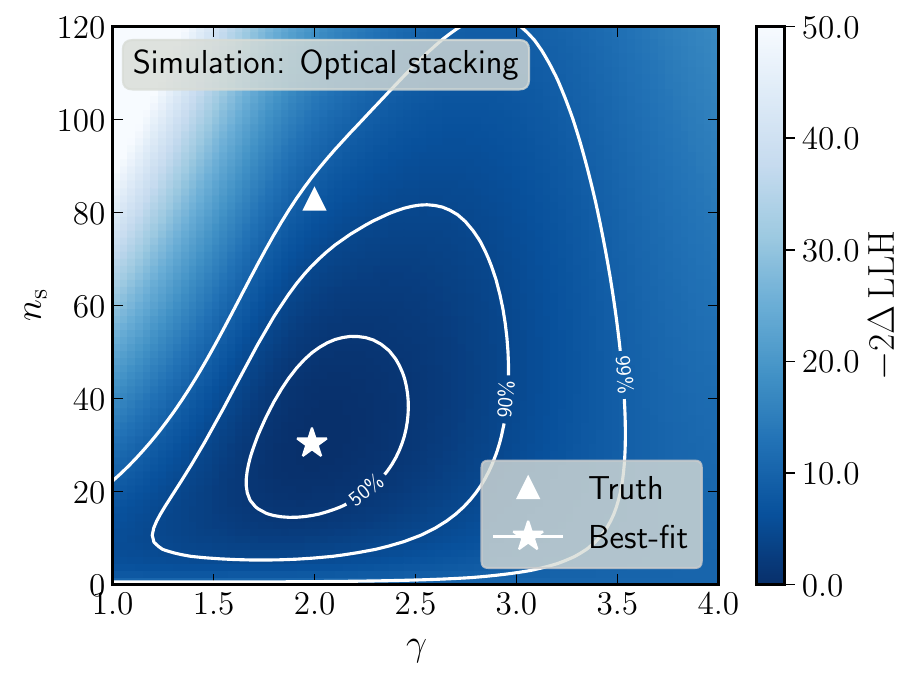}
    \caption{Likelihood contours in ($n_s$, $\gamma$) for the gamma-ray stacking analysis (left) and for the optical stacking analysis (right), reproduced from \cite{IceCube:2022lnv}. Here, $n_s$ refers to the total number of signal events for the entire stacked sample. The top row shows our observed data and the bottom row shows a sample pseudo-experiment with injected signal. Contours denote 50\%, 90\%, and 99\% containment assuming Wilk's theorem with two degrees of freedom. For our observed data (top), $\hat{\gamma}$ is undefined when $\hat{n}_s = 0$, so we do not include a best-fit point on these panels.}
    \label{fig:llh_spaces}
\end{figure*}

In addition to searching individual novae for neutrino emission, we performed two stacking analyses.  One stacked 16 gamma-ray detected novae, excluding RS Oph, because the analysis was designed prior to its 2021 outburst, to test the hypothesis that neutrino emission is proportional to the gamma-ray flux.  The other stacked $N$ optically-detected novae, to test the hypothesis that neutrino emission is proportional to the optical flux.  While gamma-ray and neutrino emission are tightly coupled at production in hadronic models, there may be substantial absorption of the gamma-ray signal, in which case the optical flux is expected to better trace the neutrino emission.  Under the particular proportionality models tested by these stacking analyses, they achieve sensitivity superior to the single-nova analyses presented in Section \ref{sec:individual}.  Results from the stacking analyses are shown in Fig.~\ref{fig:llh_spaces}.  No significant signal is found in either analysis, and we therefore place upper limits on the total number of signal neutrinos summed over all novae as a function of the power law index, for each of the two stacking hypotheses.

\section{Discussion/Conclusion}

We have presented several searches for neutrinos from novae using a new IceCube DeepCore event selection named GRECO Astronomy.  In~\cite{IceCube:2022lnv}, we described searches for neutrinos from 16 individual gamma-ray-detected novae.  That analysis was developed before the 2021 outburst of RS Oph, which was detected by MAGIC and H.E.S.S.  In addition to the real-time search for TeV-PeV neutrinos from that outburst reported in~\cite{2021ATel14851....1P}, we have now added an archival search for GeV-TeV neutrinos from RS Oph using GRECO Astronomy.  In addition to the search for neutrinos from individual novae, we performed two stacking searches, one based on gamma-ray fluxes and one based on optical fluxes.  There is no evidence for neutrino emission from novae in any of the analyses, and the upper limits we present are an order of magnitude above the measured gamma-ray fluxes and corresponding neutrino predictions.  Nevertheless, these upper limits can already reject scenarios in which a large fraction of the gamma-ray emission is absorbed within the source.

This analysis uses a new event selection, GRECO Astronomy, which is well suited for astrophysical transients and has already been used for other source classes including gravitational wave sources and gamma-ray bursts.  
The IceCube Upgrade is a fully funded project now underway to install an additional seven strings within DeepCore, with even denser instrumentation than DeepCore both horizontally and vertically.  It will directly improve the performance of IceCube in the GeV-TeV band.  In particular, event direction reconstruction will improve over that demonstrated with GRECO Astronomy, lowering the effective background rate in analyses of astrophysical sources.  In addition to the neutrino paticle physics for which it was designed, the IceCube Upgrade will enable IceCube to continue to advance neutrino astronomy in the GeV-TeV band, where gamma-ray telescopes have demonstrated the sky to be full of a variety of surprising sources including novae.


\bibliographystyle{ICRC}
\bibliography{references}

\clearpage

\input{authorlist_IceCube.tex}

\end{document}

%% file: authorlist_IceCube.tex
\section*{Full Author List: IceCube Collaboration}

\scriptsize
\noindent
R. Abbasi$^{17}$,
M. Ackermann$^{63}$,
J. Adams$^{18}$,
S. K. Agarwalla$^{40,\: 64}$,
J. A. Aguilar$^{12}$,
M. Ahlers$^{22}$,
J.M. Alameddine$^{23}$,
N. M. Amin$^{44}$,
K. Andeen$^{42}$,
G. Anton$^{26}$
% ...
\\
\\
% ...
$^{12}$ Universit{\'e} Libre de Bruxelles, Science Faculty CP230, B-1050 Brussels, Belgium \\
% ...
$^{17}$ Department of Physics, Loyola University Chicago, Chicago, IL 60660, USA \\
$^{18}$ Dept. of Physics and Astronomy, University of Canterbury, Private Bag 4800, Christchurch, New Zealand \\
% ...
$^{22}$ Niels Bohr Institute, University of Copenhagen, DK-2100 Copenhagen, Denmark \\
$^{23}$ Dept. of Physics, TU Dortmund University, D-44221 Dortmund, Germany \\
% ...
$^{26}$ Erlangen Centre for Astroparticle Physics, Friedrich-Alexander-Universit{\"a}t Erlangen-N{\"u}rnberg, D-91058 Erlangen, Germany \\
% ...
$^{40}$ Dept. of Physics and Wisconsin IceCube Particle Astrophysics Center, University of Wisconsin{\textendash}Madison, Madison, WI 53706, USA \\
% ...
$^{42}$ Department of Physics, Marquette University, Milwaukee, WI, 53201, USA \\
% ...
$^{44}$ Bartol Research Institute and Dept. of Physics and Astronomy, University of Delaware, Newark, DE 19716, USA \\
% ...
$^{63}$ Deutsches Elektronen-Synchrotron DESY, Platanenallee 6, 15738 Zeuthen, Germany  \\
$^{64}$ Institute of Physics, Sachivalaya Marg, Sainik School Post, Bhubaneswar 751005, India \\
% ...

\subsection*{Acknowledgements}

\noindent
The authors gratefully acknowledge the support from the following agencies and institutions:
USA {\textendash} U.S. National Science Foundation-Office of Polar Programs,
U.S. National Science Foundation-Physics Division,
U.S. National Science Foundation-EPSCoR,
Wisconsin Alumni Research Foundation,
Center for High Throughput Computing (CHTC) at the University of Wisconsin{\textendash}Madison,
Open Science Grid (OSG),
Advanced Cyberinfrastructure Coordination Ecosystem: Services {\&} Support (ACCESS),
Frontera computing project at the Texas Advanced Computing Center,
U.S. Department of Energy-National Energy Research Scientific Computing Center,
Particle astrophysics research computing center at the University of Maryland,
Institute for Cyber-Enabled Research at Michigan State University,
and Astroparticle physics computational facility at Marquette University;
Belgium {\textendash} Funds for Scientific Research (FRS-FNRS and FWO),
FWO Odysseus and Big Science programmes,
and Belgian Federal Science Policy Office (Belspo);
Germany {\textendash} Bundesministerium f{\"u}r Bildung und Forschung (BMBF),
Deutsche Forschungsgemeinschaft (DFG),
Helmholtz Alliance for Astroparticle Physics (HAP),
Initiative and Networking Fund of the Helmholtz Association,
Deutsches Elektronen Synchrotron (DESY),
and High Performance Computing cluster of the RWTH Aachen;
Sweden {\textendash} Swedish Research Council,
Swedish Polar Research Secretariat,
Swedish National Infrastructure for Computing (SNIC),
and Knut and Alice Wallenberg Foundation;
European Union {\textendash} EGI Advanced Computing for research;
Australia {\textendash} Australian Research Council;
Canada {\textendash} Natural Sciences and Engineering Research Council of Canada,
Calcul Qu{\'e}bec, Compute Ontario, Canada Foundation for Innovation, WestGrid, and Compute Canada;
Denmark {\textendash} Villum Fonden, Carlsberg Foundation, and European Commission;
New Zealand {\textendash} Marsden Fund;
Japan {\textendash} Japan Society for Promotion of Science (JSPS)
and Institute for Global Prominent Research (IGPR) of Chiba University;
Korea {\textendash} National Research Foundation of Korea (NRF);
Switzerland {\textendash} Swiss National Science Foundation (SNSF);
United Kingdom {\textendash} Department of Physics, University of Oxford.